\documentclass[aip,rsi,amsmath,amssymb,reprint]{revtex4-1}
\usepackage{graphicx}
\usepackage[utf8x]{inputenc}

\begin{document}

\title{A broadband and high throughput single-monochromator Raman spectrometer; application for single-wall carbon nanotubes}

\author{G\'abor F\'abi\'an}
\affiliation{Budapest University of Technology and Economics, Department of Physics, Budafoki \'{u}t 8., Budapest, H-1111, Hungary}

\author{Christian Kramberger}
\affiliation{University of Vienna, Faculty of Physics, Strudlhofgasse 4., Vienna, A-1090, Austria}

\author{Alexander Friedrich}
\affiliation{University of Vienna, Faculty of Physics, Strudlhofgasse 4., Vienna, A-1090, Austria}

\author{Ferenc Simon}
\email[Corresponding author: ]{ferenc.simon@univie.ac.at}
\affiliation{Budapest University of Technology and Economics, Department of Physics, Budafoki \'{u}t 8., Budapest, H-1111, Hungary}
\affiliation{University of Vienna, Faculty of Physics, Strudlhofgasse 4., Vienna, A-1090, Austria}

\author{Thomas Pichler}
\affiliation{University of Vienna, Faculty of Physics, Strudlhofgasse 4., Vienna, A-1090, Austria}
\date{\today}

\begin{abstract}
We present a high sensitivity single-monochromator Raman
spectrometer which allows operation with a tunable laser source. The
instrument is based on the modification of a commercial Raman
spectrometer; such instruments operate with holographic Rayleigh
filters which also act as a laser mirrors and are usually considered
as inherently narrow-band. In our design, the two tasks are
separated and the filter can be freely rotated without much effect
on the light alignment. Since rotation shifts the filter passband,
this modification allows tunable operation with efficient stray
light filtering down to 150 $\text{cm}^{-1}$. The design is
optimized for single-wall carbon nanotubes, for which the
performance is demonstrated using a tunable dye-laser source. The
spectrometer thus combines the high sensitivity with the broadband
characteristics of usual triple monochromator systems.
\end{abstract}

\maketitle

\section{Introduction}
Raman spectroscopy is an important tool in various fields of science
from biology to physics or even mineralogy. In particular, Raman
spectroscopy became the most useful method in the fundamental
studies and characterization of single-wall carbon nanotubes
(SWCNTs) \cite{DresselhausCNTRamanReview}. The success of this
method for this system and the increased demand for versatile
spectrometers lead to a boom in instrument development.

Raman spectroscopy is an inelastic light scattering method, where
the energy transfer to and from the system is characteristic to
electronic, optical, vibrational, or even magnetic properties. The
method operates with monochromatic lasers as excitation.
Spectrometers which can be used with a tunable laser source are
referred to as ``broadband'' herein. The Raman spectrum consists of the
intensity of the scattered photons (or Raman intensity) at a given
energy transfer (also called the Raman shift).

In the special case of the so-called resonance Raman spectroscopy
(RRS), the Raman intensity is enhanced when either the exciting
laser or the scattered light energy matches that of an optical
transition of the system. For instance, RRS enabled to increase 
sensitivity in heavily diluted organic compounds \cite{RRS_Biology}, 
to distinguish between different species in a minerals \cite{RRS_minearology}, 
and to selectively enhance phonon modes with particular symmetries
\cite{Falicov}. SWCNTs possess different transition
energies depending on their diameter, therefore a laser energy
dependent RRS measurement allows to identify and study SWCNTs of
different kinds.

The major challenge of Raman spectroscopy is the efficient
suppression of photons with wavelengths close to that of the
exciting laser \cite{Paillet_RSI_raman}. Such photons originate from
quasi-elastic light scattering processes such as e.g. Rayleigh
scattering and their intensity dominates over the Raman process by a
few orders of magnitude. This quasi-elastically scattered light is
often referred to as ``stray light'' or ``Rayleigh scattered light''.

The efficient stray light suppression is particularly important for
SWCNTs due to the low energy ($\geq 100-150\,\text{cm}^{-1}$)
\cite{RaoCNTRamanScience} Raman modes. This, combined with the
narrow (FWHM $\sim 30 \,\text{meV}$) optical transition energies
\cite{FantiniPRL2004} require a broadband spectrometer with
efficient stray light rejection.

Modern Raman spectrometers which allow operation down to 100 cm$^{-1}$ contain holographic Rayleigh filters rather
than the classical subtractive double monochromator system. Such spectrometers are often referred to as
\emph{notch-filter} Raman systems. The principal advantage of this development is the higher signal sensitivity as
typical transmission for the holographic filters is above $80\,\%$ for the passband, whereas transmission for the
double monochromator system is significantly lower for the frequency filtered light.

A principal drawback of Raman systems with holographic filters is their potentially narrow-band characteristics.
This is due to the fact that holographic filters are manufactured for the most common laser lines only, e.g. for those
of the Ar/Kr and He/Ne lasers. In principle, the range of filter operation could be extended by rotating them.
However for most spectrometers, the holographic filter has a double role: it reflects the laser light to the sample and it
functions as a Rayleigh filter. A rotated filter leads to a change in the optical path for the excitation that can be corrected
for with a tedious spectrometer readjustment only. This is not practical for a measurement when operation with several laser lines is required.
Then, what is gained by the higher sensitivity is lost upon the time spent with the readjustment.

Herein, we describe the modification of a commercial Raman spectrometer with holographic Rayleigh filters,
which enable broadband operation with relative ease. The improvement is based on replacing the filter with a beam splitter
which guides the laser light to the sample and transmits the emitted Raman light toward the spectrometer. The holographic filter is inserted into the path of the backscattered light behind the beam splitter where it is rotated without affecting the excitation. Thus the double role of the holographic filter is split into these two components and the rotation of the filter does not observably influence the direction of the transmitted light.
Rotation of the holographic filter influences the passband differently for the $S$ and $P$
polarizations\cite{Note1}
which is overcome by applying polarization filters on the spectrometer input. The setup operates with polarizations which are
optimized when the so-called antenna effect of SWCNTs is taken into account, i.e. that the Raman light is polarized preferentially
along the polarization of the excitation \cite{SunAntennaEffect, Jorio:PhysRevLett85:2617:(2000)}.

\section{Spectrometer setup}

Commercial Raman spectrometers are usually equipped with a built-in laser and a setup optimized for this single laser line, however the characterization of SWCNTs requires measurements with a large number of laser lines \cite{KuzmanyEPJB} or with a tunable laser system \cite{FantiniPRL2004,TelgPRL2004}. Such a high sensitivity, confocal single monochromator Raman system with a holographic Rayleigh filter can be
modified to enable broadband measurements with a tunable laser, which we demonstrate for a LabRAM commercial spectrometer (Horiba Jobin-Yvon Inc.) as an example. The key step in achieving the broadband operation was replacing the built-in holographic filter, which acts as a beam splitter and a Rayleigh filter at the same time, with a combination of a simple beam splitter and the separated holographic filter. We note herein that this modification also enables a cost effective operation with usual laser wavelengths (such as e.g. the lines of an Ar/Kr laser) since no complicated filter realignments are required.

\begin{figure}[htp]
\begin{center}
\includegraphics{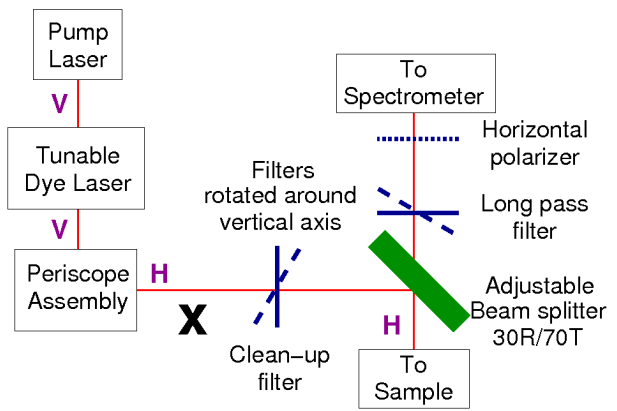}
\caption{Block diagram of the broadband configuration of the LabRAM spectrometer. V and H denote vertical and horizontal polarizations, respectively. The tunable source is a dye laser pumped with a 532 nm solid state laser. The laser light is aligned with the spectrometer using the periscope element, which also rotates the polarization from vertical to horizontal. The dye laser output is cleaned with a filter. The sample emits a nominally horizontally polarized light, which is the $P$ polarization for the long pass filter and the unwanted $S$ polarization (vertical) is filtered with the horizontal polarizer. ``X'' marks where the description is continued in the more detailed Fig. \ref{schem}.}
\label{block}
\end{center}
\end{figure}

The block diagram of the broadband spectrometer is shown in Fig. \ref{block}. The tunable excitation light source is a dye laser (Coherent Inc., CR-590) operated in the 545 nm-660 nm range with a vertical polarization. Pumping is provided by a 532 nm 5 W solid state laser (Coherent Inc., Verdi G5). Three dyes; Rhodamin 110, Rhodamin 6G, and DCM Special (Radiant Dyes GmbH) cover 545-580 nm, 580-610 nm, and 610-660 nm wavelength ranges, respectively. The periscope allows beam alignment and also changes the polarization of the light from vertical to horizontal. The dye laser output contains a spurious fluorescent background which is filtered with short pass (``3rd Millennium filters'' for 580 and 610 nm from Omega Optical Inc.) and band pass (``RazorEdge'' for 568, 633, and 647 nm from Semrock Inc.) filters \cite{Note2}. The light is directed toward the sample with a broadband beam splitter plate (Edmund Optics Inc., NT47-240) with 30 \% reflection and 70 \% transmission. A single, long pass holographic edge filter (``RazorEdge'' for 568, 633, and 647 nm from Semrock Inc.) performs stray light rejection near the laser excitation and allows operation down to $70-140\,\text{cm}^{-1}$.

\begin{figure}[htp]
\begin{center}
\includegraphics{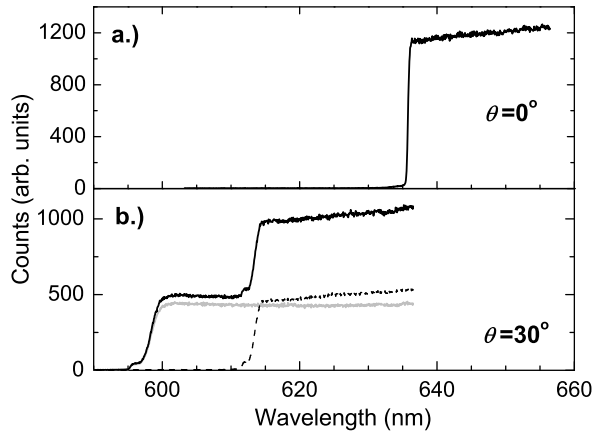}
\caption{Transmittance of the 633 nm long pass filter using unpolarized white light; a.) at normal incidence and b.) rotated by 30\textdegree. When polarization filters are used, the two parts of the double step feature (solid black line) are separated according to the $S$- and $P$-polarization (dashed black and solid gray lines, respectively). Note the broadening of filter transition width upon rotation.}
\label{LP}
\end{center}
\end{figure}

The ability to shift the transmission of the holographic filters by rotating them enables broadband operation. In Fig. \ref{LP}., we show the filter behavior at different incidence angles for a 633 nm long pass filter. The edge of transmission blue shifts upon rotation with respect to normal incidence. However, the shift is smaller for the $S$ than for the $P$ polarization; i.e. the shift is larger for the horizontally polarized light when the filter is rotated around a vertical axis, such as it is realized in our instrument.

For the $P$ polarization, the blue shift upon rotation is $\lambda(\theta) = \lambda_0 \sqrt{1 - \frac{\sin^2\theta}{n^2}}$, where $\lambda_0$ is the passband wavelength at normal incidence, $n$ is the effective index of refraction for the filter, and $\theta$ is the angle of incidence. For filters with 1 inch apertures (such as the short pass and long pass filters), rotation angles up to $30$\textdegree\ were used which yields a blue shift of about 10 \%.
The band pass ``Razor edge'' filters had an aperture of 0.5 inch which limited the blue shift to about 5 \%.

The filter transition width broadens for larger incidence angles. This is defined as the maximum difference between the laser wavelength at which the attenuation exceeds optical density 6 and the filter edge-wavelength at the 50 \% transmission point. For the 30\textdegree\ incidence, a fivefold increase in the transition width is observed when compared to the normal incidence. For the "RazorEdge" filters, the $0.2 \% \times \lambda$ transition width at normal incidence broadens to $1\% \times \lambda \approx 150 \,\text{cm}^{-1}$, which makes it usable for Raman shifts above $\sim 150 \,\text{cm}^{-1}$. The less shifted $S$ (in our construction vertically) polarized stray light is removed with a polarization filter before the spectrometer input.

\begin{figure}[htp]
\begin{center}
\includegraphics{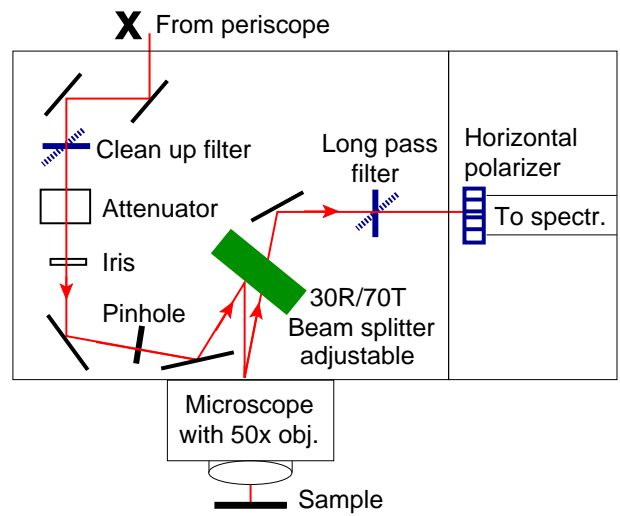}
\caption{The optical configuration of the modified LabRAM spectrometer including the location for the laser and long pass filters, the beam splitter, and the polarizer. ``X'' marks the position which is also shown in the block diagram in Fig. \ref{block}.}
\label{schem}
\end{center}
\end{figure}

Fig. \ref{schem}. shows the setup for the modified LabRAM spectrometer. For the original spectrometer, the long pass filter mirrors the light toward the sample and also acts as a Rayleigh filter. In our construction, the beam splitter mirrors the light toward the sample and the stray light is filtered with the separate long pass filter, therefore the double function for the filter is split. The broadband beam splitter plate has 30 \% reflection and 70 \% transmission, thus only a small fraction of the Raman light is lost. The 70 \% excitation power loss on the beam splitter can be compensated by reducing the attenuation of the intensive laser beam, maintaining a constant irradiation density on the sample. The beam splitter plate is mounted on a finely adjustable 2-axis holder (Thorlabs Inc., VM1) with a home made mounting. The fine adjustment is required to set the light alignment properly with the spectrometer. Final fine adjustment is performed with the holder to maximize the Raman signal.

The long pass filter is rotated around a vertical axis, which is more practical than rotation around a horizontal one. This preferred axis fixes the optimal light polarization for the whole setup. The vertical rotation axis prefers a horizontally polarized scattered (Raman) light as it is of the $P$ polarization, for which the edge shift is larger. For SWCNTs, the Raman light is polarized preferentially along the polarization of the excitation, a phenomenon called antenna-effect \cite{SunAntennaEffect, Jorio:PhysRevLett85:2617:(2000)}. Therefore a horizontally polarized laser excitation is preferred which explains the polarizations used in our design. We also verified that the LabRAM spectrometer itself is not polarization-sensitive in contrast to an older triple monochromator system (Dilor XY triple monochromator system) which is about 3 times more sensitive to the vertical polarization of the scattered light than for the horizontal one.

\section{Measurements and discussion}

Measurements were carried out on a HiPCO sample (Carbon Nanotechnologies Inc., Houston, Texas), which was suspended in a 2 weight\% SDBS (Sodium dodecyl benzene sulfonate) in water using sonication.
We focused on the radial breathing mode (RBM) Raman range located below 400 $\text{cm}^{-1}$, which is commonly studied to characterize the diameter distribution in SWCNTs \cite{DresselhausTubes}. Carbon-tetrachloride was used for Raman shift correction and Raman intensity normalization such as in Ref. \onlinecite{FantiniPRL2004}. The LabRAM spectrometer allows to change between a macro and micro mode with a mirror. The macro mode is fitted with a sample holder for cuvettes, that was provided with the spectrometer. The suspended HiPCO sample was placed in a glass cuvette under the objective of the built-in microscope (Olympus LMPlan 50x/0.50, inf./0/NN26.5, yields about $1 \times 1 \,\mu \text{m}^2$ spot size) and the CCl$_4$ reference sample was placed into the macro cuvette holder. This allowed a stable and robust spectral shift calibration and intensity normalization since the macro/micro mode change is reproducible and there is no need for further adjustments nor for sample exchange.

29 Raman spectra were acquired for laser excitation energies between 1.92 eV (648 nm) and 2.27 eV (545 nm) with an energy resolution of about 12 meV ($\sim 100\,\text{cm}^{-1}$). The spectrometer was operated with a 600 grooves/mm grating and a liquid nitrogen cooled CCD with 1024 pixels along the spectral direction, which yields a $\sim 1.3$ cm$^{-1}$ Raman shift resolution and $\sim 1800$ cm$^{-1}$ spectral range for 600 nm (both are wavelength dependent). Typical laser powers of 1-5 mW were used with no observable heating effects. This is due to the liquid nature of both samples as a bucky-paper HiPCO sample is heated observably for powers above 1 mW.

\begin{figure}[htp]
\begin{center}
\includegraphics{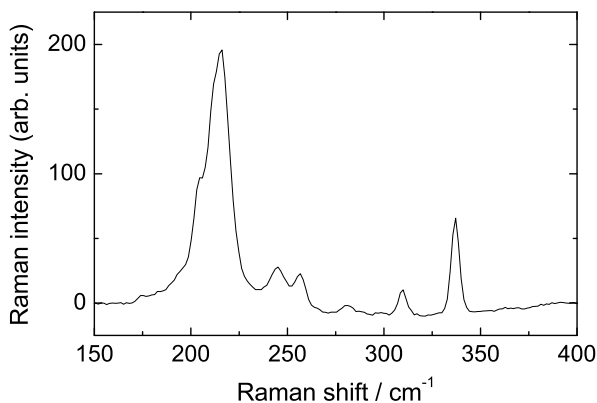}
\caption{Raman spectrum of the HiPCO/SDBS suspension for excitation at 592 nm (2.1 eV) with the broadband spectrometer for the RBM range.}
\label{spectrum}
\end{center}
\end{figure}

Typical measurement times of 4 minutes for the sample and a few seconds for the reference yields an acceptable signal to noise ratio of about $300$ as it is shown in Fig. \ref{spectrum}. The sample and reference measurements followed each other immediately without moving the grating position, which led to an accurate Raman shift measurement. A measurement cycle which consists of changing the dye laser wavelength, rotating of the laser clean-up and the long pass filters to the appropriate positions, shifting the spectrometer grating, nulling the spectrometer, and acquiring
the spectra, lasted typically 8 minutes. Additional time is needed for the filter exchange (a few minutes) and to change the laser dye and to readjust the beam alignment (about 1 hour). We note that once the dye laser is set and the light path is properly aligned with the spectrometer, no further realignment is required when the wavelength is changed, even though the filters are rotated and the light is minutely displaced.

In total, 9 hours were required to complete the energy dependent Raman experiments with the 29 laser lines including 2 laser dye exchanges. Measurements of a similar scale such as published previously using a triple monochromator spectrometer \cite{SimonPRB2006} last for about 2 weeks, mainly due to the approximately 50 times smaller S/N ratio and the need for a spectrometer realignment upon wavelength change.

\begin{figure}[htp]
\begin{center}
\includegraphics{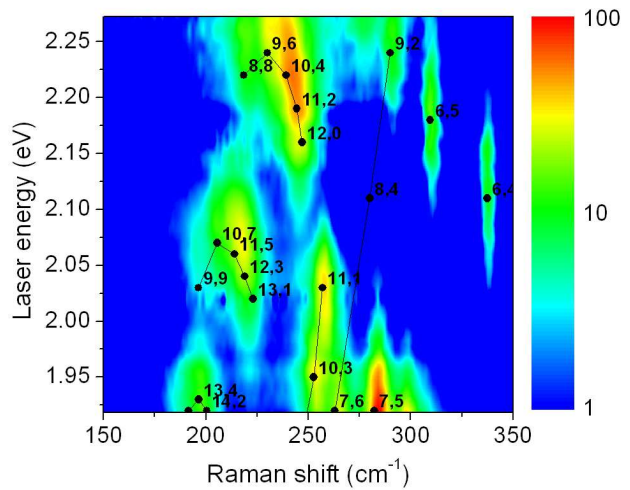}
\caption{Raman map of HiPCO/SDBS suspension measured with the broadband Raman setup.\newline Logarithmic color scale shows the Raman intensity normalized to the maximum observed intensity. Full circles denote data published in Ref. \onlinecite{FantiniPRL2004}.}
\label{map}
\end{center}
\end{figure}

Raman shift correction, intensity normalization, and the Raman map preparation was performed using a home-made software. Linear baseline correction was sufficient since no fluorescent response was encountered. Raman shift was corrected for with the carbon-tetrachloride Raman modes at 218 and 314 cm$^{-1}$. Intensity normalization using CCl$_4$ is required as it accounts for instrumental uncertainties such as a slight misalignment of the scattered light upon dye exchange. We checked the consistency of the normalization by measuring the same wavelength with different laser dyes after spectrometer realignment.

Fig. \ref{map}. shows the 2D contour plot of the Raman map. The normalized Raman intensity is displayed on a logarithmic scale. A good agreement is observed between our data and the measurements in Ref. \onlinecite{FantiniPRL2004}, whose resonance transition energies and Raman shifts are shown for the different SWCNT chiral indexes, $(n,m)$, with full circles. We do not observe the (8,4) SWCNT in our measurement. Points corresponding to the same SWCNT families, i.e. when $2n+m$ is constant \cite{DresselhausTubes}, are connected by solid lines.

\begin{figure}[htp]
\begin{center}
\includegraphics{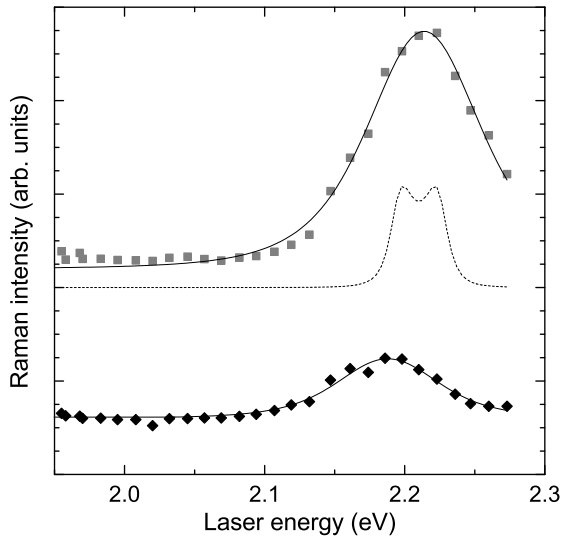}
\caption{Energy cross section of the Raman map for the RBM of two SWCNTs. The gray squares and black diamonds correspond to cross sections at Raman shifts of $244$ cm$^{-1}$ and $310$ cm$^{-1}$, respectively. Solid curves refer to the resonance Raman fits as given in the text. Dashed curve shows the calculated resonance Raman profile with a smaller quasiparticle scattering rate, $\Gamma=10\,\text{meV}$. The curves are vertically offset for clarity.}
\label{cross}
\end{center}
\end{figure}

The agreement shows the utility of the broadband arrangement with a clear advantage over previous results in terms of acquisition time. Vertical, i.e. energy cross section of the Raman map were obtained by averaging around a given Raman shift. Representative energy cross section data are shown in Fig. \ref{cross}., along with fits using the resonance Raman theory \cite{FantiniPRL2004,SimonPRB2006}:
\begin{gather}
I(E_l) \propto
\left|\frac{1}{(E_l - E_{22} - i\Gamma)(E_l \pm E_{ph} - E_{22} - i\Gamma)}\right|^2,
\label{RRformula}
\end{gather}
\noindent where $I(E_l)$ is the laser energy ($E_l$) dependent Raman scattering intensity, $\Gamma$ is the quasiparticle scattering rate, $E_{22}$ is the transition energy, $E_{ph}$ is the phonon energy.

Fits yield transition energies and quasiparticle scattering rates of $E_{22} = 2.199(1)$ eV and $\Gamma = 71(2)$ meV for the (11,2) SWCNT with $\nu_{\text{RBM}}=244$ cm$^{-1}$, and $E_{22} = 2.169(2)$ eV and $\Gamma = 59(3)$ meV for the (6,5) SWCNT with $\nu_{\text{RBM}}=310$ cm$^{-1}$. These values agree well with typical literature values when it is taken into account that the different solvent environment slightly modifies the Raman transition energies \cite{FantiniPRL2004}.

We note that no further corrections were made to obtain the energy cross section data apart from the normalization by the reference. The result is therefore remarkably smooth in comparison with similar data published in Refs. \onlinecite{FantiniPRL2004,TelgPRL2004,DoornPRB2008}. This is due to the robust and reproducible measurement of the reference sample and possibility of measuring Raman spectra at different wavelengths without spectrometer readjustment in between.

\section{Conclusions}

In conclusion, we presented the broadband modification of a high sensitivity commercial Raman spectrometer. The improvement allows the use of a tunable dye laser. The spectrometer performance is demonstrated on SWCNTs where such tunable measurements are inevitable to obtain meaningful insights into the vibrational and electronic properties.

\begin{acknowledgments}
Work supported by the Austrian Science Funds (FWF) project Nr. P21333-N20, by the European Research Council Grant Nr. ERC-259374-Sylo, and by the New Hungary Development Plan Nr. T\'{A}MOP-4.2.1/B-09/1/KMR-2010-0002.
\end{acknowledgments}

\end{document}